\documentstyle[psfig]{mn}

\input{epsf}

\def\lsim{\mathrel{\hbox{\rlap{\hbox{\lower4pt\hbox{$\sim$}}}\hbox{$<$}}}}
\def\gsim{\mathrel{\hbox{\rlap{\hbox{\lower4pt\hbox{$\sim$}}}\hbox{$>$}}}}

\def\and   {\rm {et al.} \rm}  
\def\etal  {\rm {et al.} \rm}

\begin{document}

\title[Anisotropies in the redshift-space correlations I: Simulated catalogues]
{Anisotropies in the redshift-space correlations of galaxy groups and clusters I: Simulated catalogues }

\author[N. D. Padilla \& D. G. Lambas]
{N. D. Padilla$^{1}$ \& D. G. Lambas$^{2,3}$ \\
1.University of Durham, South Road, Durham DH1 3LE, UK\\
2.IATE, Observatorio Astron\'omico de C\'ordoba, Laprida 854, 5000, C\'ordoba,
Argentina\\
3.John Simon Guggenheim Fellow\\
\\
{\rm High resolution plots and info at http://star-www.dur.ac.uk/$\sim$nelsonp/anisotropies}\\
}

\maketitle

\begin{abstract}
We analyse 
the correlation function of mock galaxy clusters in redshift space.
We constructed several mock catalogues designed to mimic the selection 
biases inherent in a variety of observational surveys.  We explore
different effects that contribute to the distortion of the clustering pattern;  
the pairwise velocity distribution of galaxy systems, coherent 
bulk motions, redshift errors and systematics in cluster identification.
Our tests show that the redshift-space clustering pattern of galaxy 
systems is highly influenced by effects associated with the identification
procedure from two dimensional surveys.  These systems show a spuriously 
large correlation
amplitude, an effect that is present and even stronger in  a subsample 
whose angular positions coincide with 3-dimensional identified clusters.
The effect of a small  number of
redshift measurements is also that of increasing the correlation amplitude.
In a similar fashion, the bias parameter
inferred from cluster samples is subject to these observational problems
which induce variations of up to a factor of two in such determinations.
Also, we find that
the estimated mean pairwise velocity dispersion can be up to an order of 
magnitude larger than the actual value.
Errors in the estimated cluster redshift, originating from the use of too few
redshift measurements per cluster, have a smaller impact on the 
measured correlation function.  We show that 
an angular incompleteness in redshift surveys,
 such as that present in the 2dFGRS 100k public 
release, has no significant effect in the results.  We suggest that
the nature of projection effects arise mainly from structures along the
line of sight in the filamentary large-scale clustering pattern.  Thus,
spectroscopic surveys are the only means of  providing 
unbiased cluster samples.
\end{abstract}


\section{Introduction}

Rich clusters of galaxies form at the highest 
peaks of the mass distribution
through physical processes which are simpler than those
involved in galaxy formation.  As a consequence, models show that
the correlation function and 
power spectrum of clusters of galaxies have the same shape
as that of the mass, with an amplitude depending
on a small number of parameters (see for example Colberg et al. 2000,
Sheth, Mo \& Tormen 2001,
Padilla \& Baugh, 2002). In fact, the space density of clusters of galaxies 
is usually the only parameter needed to obtain a direct estimate of the
underlying mass distribution (Padilla \& Baugh, 2002).  
This shows the clear advantage
of studying clusters of galaxies, which has made them
popular tracers of the large scale structure of the Universe
in the literature.  

Cluster distances are 
derived from member galaxy redshifts, which 
are affected by galaxy peculiar velocities.  
Even if all the member galaxy redshifts are used in the determination
of the cluster distance, this will be still affected by the cluster
peculiar motion, and will not be a true distance measurement.
The distribution function of cluster peculiar motions 
depends upon the value of the mass density parameter $\Omega$, for models 
with similar mass fluctuation amplitudes (Croft \& Efstathiou 1994a; 
Bahcall, Cen \& Gramann 1994). These motions, either
for galaxies or clusters of galaxies, produce an apparent
distortion of the clustering pattern as measured by the
two-point correlation function in redshift space, $\xi(\sigma,\pi)$,
where $\sigma$ and $\pi$ are the separations perpendicular and
parallel to the line of sight, respectively.
At small separations, non-linear,  nearly virialized regions produce
elongations along the line of sight which allow measurements 
of the one-dimensional pairwise rms velocity dispersion, $<w_{12}^2>^{1/2}$
(Davis \& Peebles 1983).  Larger scales are dominated
by the infall onto overdense regions in the form of bulk motions 
which  result in a compression of the $\xi$ contours
along the direction of the line of sight.

Evidence of 
strong anisotropies in the correlation function of
Abell clusters has been given by several authors (see for instance 
Postman, Huchra \& Geller 1992 and references therein).
Systematic effects in the Abell catalogue originate in the superposition of 
clusters along the line of sight 
and generate  a bias in the observed correlation function of the catalogue 
(Sutherland 1988; Sutherland \& Efstathiou 1991; see also Lucey 1983).
Using mock catalogues from numerical simulations,
van Haarlem \etal (1997) showed that a large fraction of Abell clusters of
richness class $R \geq 1$ would not be 
physically bound systems and would suffer from contamination
by galaxies and groups along the line of sight.  
Therefore it is likely that the observed
large anisotropies could be mainly produced by systematics
in cluster detection algorithms
from angular galaxy catalogues.

The construction of objectively defined cluster 
catalogues drawn from machine-scanned survey plates with better 
calibrated photometry brought a new generation
of cluster catalogues which are believed to be less affected
by identification biases (APM: Dalton \etal 1992, 1994, 1997; 
Cosmos: Lumsden \etal 1992). 
The typical radius used to define clusters in the machine 
based catalogues is significantly smaller than that used by Abell, 
reducing the enhancement of cluster richness by projection effects. 
The clustering signal found in these more recent cluster redshift surveys 
does not display large enhancements along the line of sight; furthermore, 
the trend of increasing correlation amplitude with decreasing space 
density of clusters is weaker than that found for Abell clusters 
(Croft \etal 1997) and is similar to the trend expected in current models
of structure formation. A similarly weak dependence of correlation 
length on cluster space density is found in X-ray selected cluster 
catalogues, such as the X-ray Bright Abell Cluster Sample (XBACS) and 
the REFLEX sample (Collins et al. 2000), 
which are less susceptible to line of sight 
projection effects than the optically selected  Abell catalogue 
(Abadi, Lambas \& Muriel 1998; Borgani et al. 1999).
However, Miller \etal (1999) argue  
that the early redshift surveys of Abell clusters 
contain large fractions of low richness clusters 
(Abell richness class $R=0$), which were not intended to form complete samples 
suitable for statistical analyses. 
Miller \etal (1999) present the clustering analysis of a new redshift 
survey of Abell clusters with richness $R \ge 1$, and with the majority of 
cluster positions determined using several galaxy redshifts. 
The clustering signal along the line of sight is greatly reduced in the 
new redshift surveys compared with the Bahcall \& Soneira (1983) 
results, and is comparable to the amount of distortion of the 
clustering pattern found for APM clusters (see Fig 5 of Miller \etal 1999).
The anisotropy is further reduced after the orientation of two 
superclusters that are elongated along the line of 
sight is changed. 
Peacock \& West (1992) also found that restricting the attention to higher 
richness Abell clusters removed the strong radial anisotropy seen 
in the clustering measured in the earlier surveys.

Another possible source of systematics could rely
on the fact that several cluster distances are
determined by a single galaxy 
redshift, usually the brightest cluster member. 
Therefore it is important to explore the effects of  
cluster distance errors 
on the analysis of correlation function of rich clusters.  

The gravitational amplification of small primordial fluctuations 
has been analysed through numerical simulations to explore the spatial 
distribution of clusters 
(e.g. White \etal 1987; Bahcall \& Cen 1992; Croft \& Efstathiou 1994b; 
Watanabe, Matsubara \& Suto 1994; Eke \etal 1996).  
These early studies do not reach a consensus on the predicted 
clustering of clusters in cold dark matter cosmologies. 
Part of the reason for this discrepancy is due to 
differences in the way in which 
clusters are identified in the simulations (Eke \etal 1996). 
More recent studies have made use of much larger volumes than 
in these earlier studies, with sufficient resolution to 
allow the reliable extraction of massive dark matter haloes that 
can be identified as rich clusters (Governato \etal 1999; 
Colberg \etal 2000, Jenkins \etal 2001).

In this paper, we analyse the redshift space clustering of massive 
dark matter haloes in mock catalogues extracted from
the $\tau$CDM Hubble Volume simulation.  In particular,
we use the mock catalogues made available by the Virgo Consortium, which
contain galaxy angular positions and distances, with a redshift distribution
set by a selection function.
In this work we will make use of these mock catalogues to
analyse in detail the effects of projection biases on catalogues
of clusters identified in two dimensions.  We also predict
the results of the new generations of cluster catalogues with a
high degree of spectroscopic completeness such as the 2dF Galaxy
Redshift Survey (Norberg et al. 2002).

The outline of this paper is as follows: section 2 describes the equations
which allow us to obtain a statistical description of the redshift-space
correlation function as a function of coordinates 
parallel and perpendicular to the
line of sight, through the pairwise velocity dispersions, redshift-space 
correlation length and bias parameters; 
section 3 presents the correlation function measurement technique.
Section 4 presents
the different mock cluster samples constructed using several cluster
identification algorithms, and studies the results arising from using different 
number of redshifts to derive cluster distances 
and also different values of the 
search radius in cluster identification.  In this section,
we also study the dependence of the relative pairwise velocities,
redshift-space correlation length and bias factors as a function
of the number of redshifts used in the determination of the cluster distances.
In section 5 we discuss our results and present the main conclusions 
drawn from this work.

\section{Redshift-space correlation
function anisotropy}

The redshift-space correlation function can be calculated as a function of
the pair separation parallel and perpendicular to the line of sight.  This
approach has led to the quantification of characteristics of the
redshift-space distribution of galaxy and clusters of galaxies, such as 
the "fingers of God", which are elongated structures seen in redshift
surveys, originated from the random motions of galaxies inside clusters.  

The strength of the ``fingers of god" effect depends on  the 
pairwise velocities of galaxies, which so
far has been measured using two different methods.
The first method corresponds to that adopted by 
Loveday et al. (1996) and Ratcliffe et al. (1998), which compares
a theoretical expression for the correlation function with the measured
values, $\xi^m(\sigma,\pi)$ (the index $m$ indicates measured quantities),
in a grid of values of distances parallel and perpendicular
to the line of sight.  The other approach presented by
Padilla \etal (2001), compared the isopleths (curves of equal
correlation function amplitude) in the measured
and predicted correlation functions.
The $\xi(\sigma,\pi)$  contour levels are approximated
by the functions $r^m(\theta)$ and $r^p(\theta)$ for the
measured and predicted correlations respectively.
Here $\theta$ is the polar angle measured from the direction perpendicular 
to the line of sight (Padilla et al. 2001), such that 
$\xi^{x}(\sigma_r,\pi_r)-\xi^{fix}=0$,
where $\sigma_r=r^{x}_{\xi^{fix}}(\theta)\cos(\theta)$, 
$\pi_r=r^{x}_{\xi^{fix}}(\theta)\sin(\theta)$, and
the index $x$ indicates either measured ($m$) or predicted
quantities ($p$).

In either method, the measured correlation function  $\xi^m(\sigma,\pi)$ 
is compared with 
the convolution of the real-space correlation function, $\xi(r)$ 
with the pairwise velocity distribution function, $f(w)$, following
Bean et al. (1983).  We calculate
\begin{equation}
1+\xi^p(\sigma,\pi)=\int_{-\infty}^{\infty} [1+\xi(r)] f[w' - w^s(r,r')]{\rm d}w',
\label{eq:pred}
\end{equation}
where $r^2=r'^2+\sigma^2$, and $H_0$ is the Hubble constant, $r'=\pi - w'/H_0$ 
(the prime denotes the line-of-sight component of a vector quantity) and 
\begin{equation}
w^s(r,r') \simeq -H_0 \beta \xi(r) (1+\xi(r))^{-1}r',
\label{eq:streaming}
\end{equation}
is the mean streaming velocity of galaxies at separation $r$.  

Following usual procedures we calculate 
the best-fit rms peculiar velocity,
$<w^2>^{1/2}$, for an exponential distribution, 
\begin{equation}
f(w)=\frac{1}{\sqrt{2}<w^2>^{1/2}}\exp \left( -\sqrt{2} \frac{|w|}{<w^2>^{1/2}} \right).
\label{eq:fw}
\end{equation}
We adopted this pairwise velocity distribution as it has shown 
to be the most accurate fit to the results from numerical simulations
(Ratcliffe et al. 1998).

\begin{figure*}
{\epsfxsize=16.truecm \epsfysize=8.truecm 
\epsfbox[33 313 567 561]{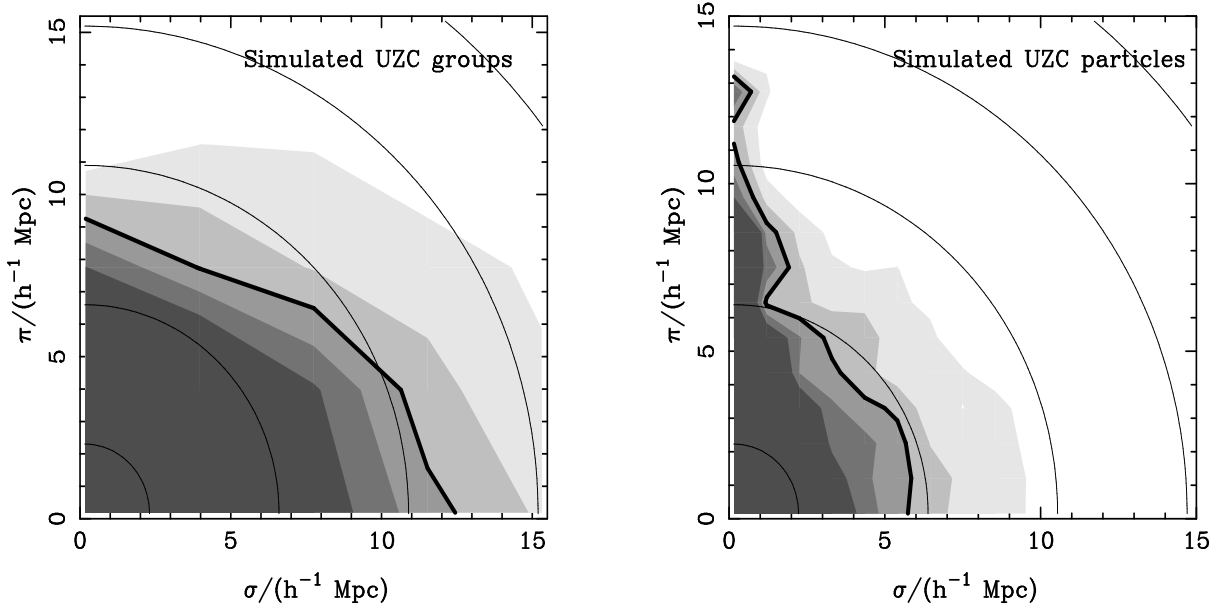}}
\caption{
The 2-point 
correlation function of simulation groups (left hand side panel) and 
simulation particles (right panel) in the coordinates $\sigma$ and $\pi$
(mock group and galaxy samples from Padilla et al., 2001). 
From light to dark, the transitions between different 
shadings correspond to fixed level values of $\xi=0.6,0.8,1.0,1.2,$ and $1.4$,
which are later used to infer the relative pairwise
velocities, $w_{12}$. 
The thick line corresponds to the $\xi=1$ level, and the thin lines
show the expected contours in the absence of peculiar velocities. 
}
\label{fig:fig6_distclusters}
\end{figure*}
Two estimates of the
real-space correlation function are usually adopted
in equation \ref{eq:pred}:
the inversion of  the projected correlation
function (Baugh 1996, Ratcliffe et al. 1998), 
or a simple power-law fit
obtained from the angular correlation function (Padilla et al. 2001).
In this work we use the theoretical 
correlation function of the simulated mass distribution 
and adopt a linear bias parameter $b$ that relates clusters and mass 
by Fourier transforming the model power spectrum 
\begin{equation}
\xi^{CDM}(r)= \frac{1}{2\pi^2} b^2 \int_0^{\infty} P(k) 
\frac{\sin(kr)}{kr} k^2 {\rm d}k,
\label{eq:xicdm}
\end{equation}
We use this to evaluate the theoretical prediction for $\xi^p(\sigma,\pi)$
using equation (\ref{eq:pred}).  We 
search for the optimum values of the scale independent bias parameter 
$b$ and $<w^2>^{1/2}$ by minimising the quantity $\chi^2$,
\begin{equation} 
\chi^2 = \sum_{i} [r^m_l(\theta_i)-r^p_l(\theta_i)]^2,
\label{eq:chi}
\end{equation} 
where we have chosen to compare a set of discrete levels, $l=0.6,0.8,1.0,1.2$
and $1.4$, of the redshift-space 
correlation function $\xi(\sigma,\pi)$ amplitude instead
of comparing values of the correlation function on a grid of $\sigma$ and 
$\pi$ distances.  Our
choice is based on the fact that more reliable and stable results are 
obtained using this technique (Padilla et al. 2001).

\section{Measuring $\xi(\sigma,\pi)$}
\label{ssec:measuringxi}

We have computed 
$\xi^m(\sigma,\pi)$
as a function of the separation
perpendicular ($\sigma$) and parallel ($\pi$) to the line of sight.
To compute $\xi^m(\sigma, \pi)$, we generate a random catalogue with the 
same angular limits and radial selection function as the sample of objects.
We cross correlate data-data and random-random pairs ($N_{dd}$ and $N_{rr}$ 
respectively) binning them as a function
of separation in the two variables $\sigma$ and $\pi$.  Our estimate 
of $\xi^o(\sigma,\pi)$ is (Davis \& Peebles 1983):
\begin{equation}
\xi^m(\sigma,\pi)=\frac{N_{dd} n_R^2}{N_{rr} n_D^2}-1
\end{equation}
where $n_D$ and $n_R$ are the number of data and random points respectively. 

This estimator is affected by uncertainties in the mean density,
in particular on large scales where $\xi(r)$ is small (eg. Hamilton 1993). 
However, since our analysis is 
confined to small separations where the correlation function has
a large amplitude, our results are insensitive to the choice of estimator.

Figure \ref{fig:fig6_distclusters} shows the contours of equal amplitude of 
the 2-point correlation function of groups identified
in a numerical simulation (left panel) and simulation particles
(right panel) in the coordinates $\sigma$ and $\pi$ (mock
galaxy and group samples from Padilla et al. 2001).  The compression of the 
iso-correlation curves for groups in the $\pi$ direction can be appreciated,
indicating a lack of 
high pairwise velocities in this sample.  From the comparison with results 
derived from clusters identified in two-dimensional surveys (see for instance
Bahcall, Soneira \& Burgett, 1986) 
the existence of either large pairwise 
velocities or large projection biases is clear, as also discussed by Sutherland (1988).

This elongation could originate in 
the systematic presence of groups along the line of sight, in the fields
of clusters identified from angular data.
From the theoretical point of view, such strong elongations along the line of
sight are not expected in a hierarchical scenario of structure formation.
This is confirmed by the compression observed in left hand side panel
of figure \ref{fig:fig6_distclusters} and 
provides a clear evidence for the infall of groups onto larger structures
(see also Peacock et al. 2001, for galaxies). 
As the contours are well defined for all the correlation function 
levels shown in this figure ($\xi=0.6$, $0.8$, $1.0$,
$1.2$ and $1.6$), the analysis 
developed in the next sections will be based on the results at these
contours.
In the right hand side panel, the large distortion seen in the 
iso-correlation curves in the $\pi$ direction   
is due to the larger peculiar velocities of the simulation particles
and their smaller correlation amplitude, associated with the 
non-linear evolution of the density field on small scales.

\section{Results from mock catalogues extracted from N-body simulations}
\label{s:nbody}

In this section, we analyse the statistical properties of cluster 
samples drawn from mock catalogues.  These mock
catalogues have been obtained from one of the mock galaxy
catalogues extracted from the $\tau$CDM Hubble Volume
simulation by the Durham Extragalactic Group (Evrard et al., 2002)
following a procedure similar to that used in Cole et al. (1998). 
This simulation follows the evolution of cold dark matter (CDM) 
density fluctuations in a $\tau$CDM cosmology, with parameters 
$\Omega_{0} = 1.0$, a cosmological constant 
$\Lambda_{0}c^{2}/(3H_{0}^{2}) = 0$, and a power spectrum described 
by a shape parameter of $\Gamma = 0.21$ 
and $\sigma_{8} = 0.6$. The huge volume of the simulation 
($8h^{-3} {\rm Gpc}^{3}$)
and the large number of particles employed ($10^{9}$) 
allow cluster statistics to be studied with unprecedented accuracy 
(Colberg \etal 2000; Jenkins \etal 2001).

The particular mock catalogue we chose to apply the different
identification methods to, is that of the APM survey, with
a selection function consistent with a limiting magnitude
$b_J = 20.0$.

This particular mock catalogue has the advantage
that the derived galaxy apparent magnitudes have $k+e$ corrections 
which reproduce 
the results of the 2dFGRS (see Norberg et al. 2002, for details).
Also we notice that the observer is not chosen 
at random but instead constrained to lie 
in a region with similar properties to those of the Local Group.  

\subsection{Mock Groups identification}

\begin{figure*}
{\epsfxsize=15.truecm \epsfysize=9.truecm 
\epsfbox[ 1 1 765 415]{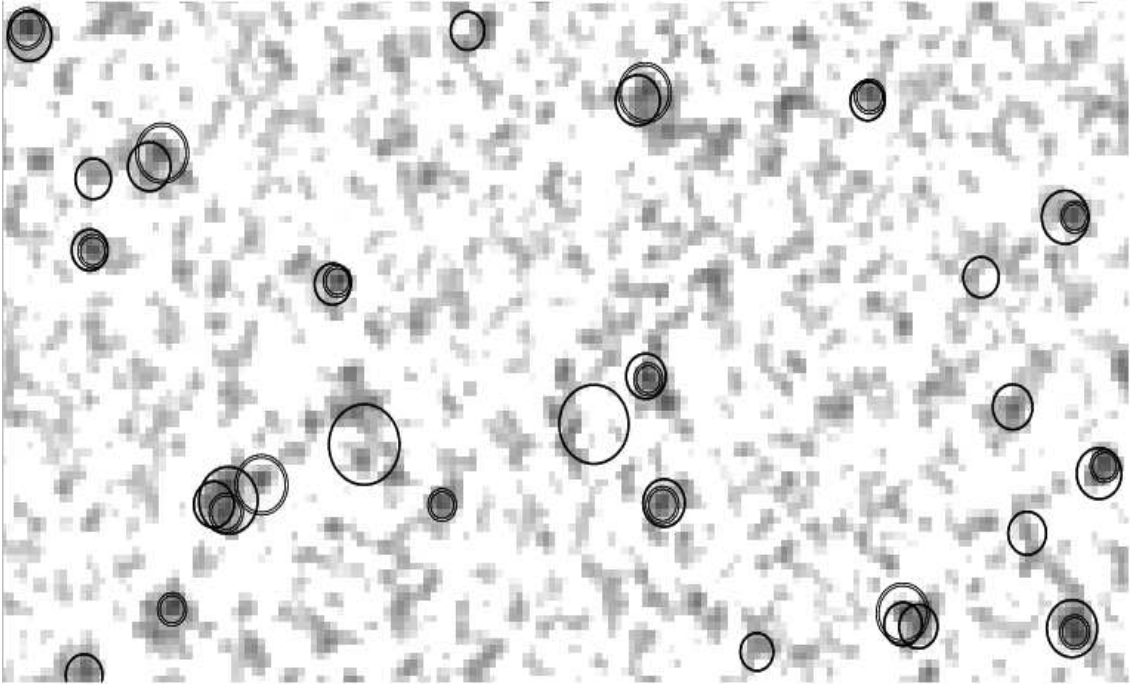}}
\caption{
Galaxy density and clusters in the mock catalogues.  The density
of galaxies in the pixelmap is proportional to the darkness of the
pixels in this figure.  The double-line circles show clusters
from mock sample 3, whereas thick-line circles indicate the positions
of clusters in mock sample 2.  This figure shows a sub-set of the
pixelised mock galaxy catalogue and is shown 
in an equal area projection centred at
$\alpha=1^h$ and $\delta=-45^o$, with radius $r_{eqa}=160^o$.  The
number of pixels per side is $n_{pix}=2900$, and was set in order
to obtain a mean number of galaxies $n=1$ per pixel.  The pixelmap sub-set
size is $170{\rm x}90$ pixels.
}
\label{fig:pixel_2dconf}
\end{figure*}

\begin{figure*}
{\epsfxsize=15.truecm \epsfysize=15.truecm 
\epsfbox[-31 22 567 561]{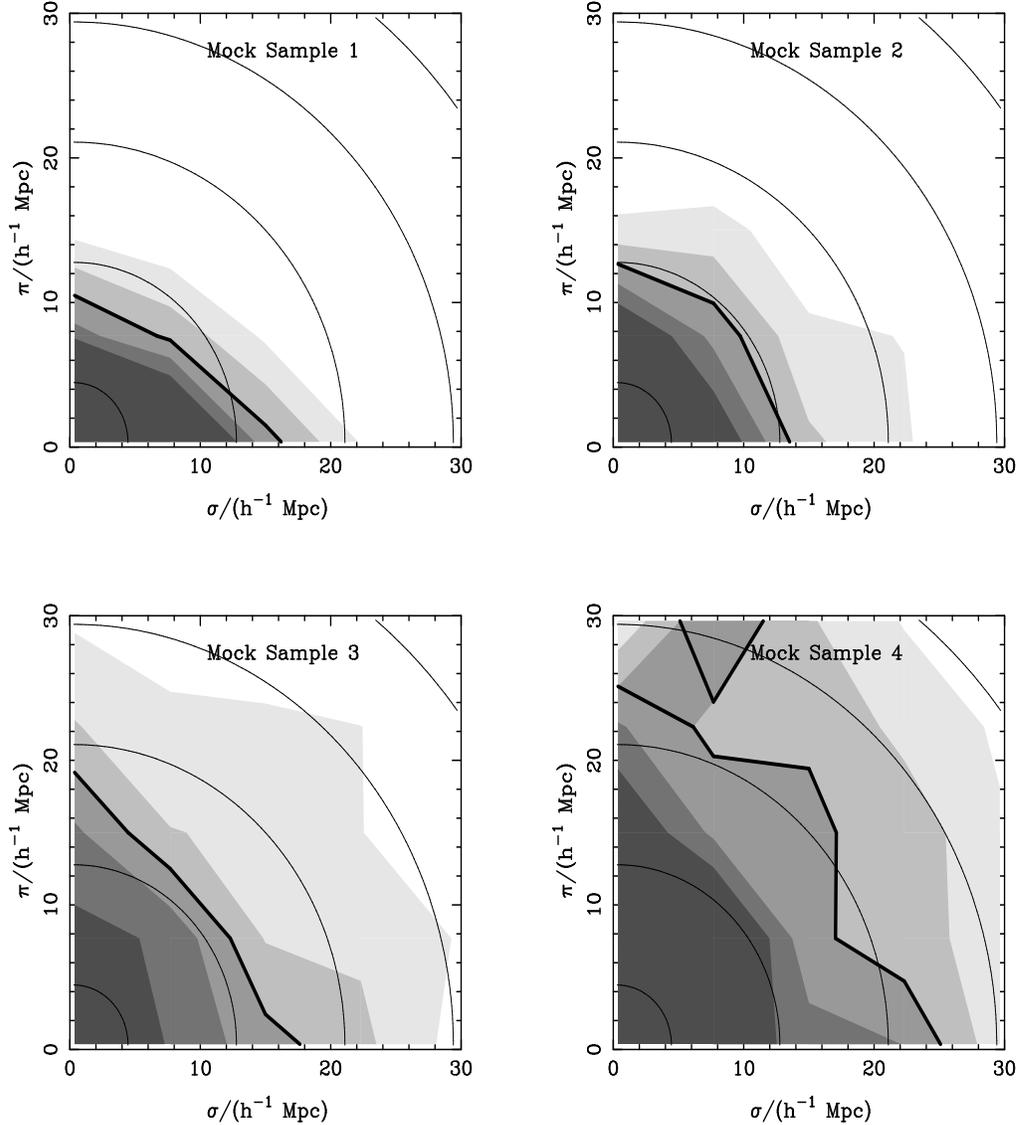}}
\caption{
$\xi(\sigma,\pi)$ from simulated catalogues of clusters of galaxies
with distances measured using $10$ member galaxies, obtained
using different algorithms.  Upper left panel shows
the results from sample 1, upper right panel from sample 2,
lower left panel shows the contours obtained from sample 3,
and the lower right panel, those from sample 4.
Shadings and line conventions are as in figure 1.
}
\label{fig:dist.sim}
\end{figure*}

We use three main algorithms to search for groups in 
the numerical simulations,
and construct four samples of clusters from them:
\begin{itemize}
\item Sample 1, is constructed using a 
friends-of-friends algorithm  applied to
the 3-dimensional distribution of particles in the simulation cube.
We used a linking length $b=0.2$, expressed in terms
of the mean interparticle separation, adequate for the
$\tau$CDM cosmology (Helly et al., 2003).  
This sample is almost completely free of spurious clusters since
we have chosen to study only those groups with at least $60$ member 
particles.  The fraction of unbound associations found using FOF
drops quickly as the number of member particles rises over
$10$.

Given the lack of a significant number of 
spurious clusters, the measurements
of anisotropies in the correlation function should clearly show
the expected infall pattern.  As we do not have computational 
access to the full  Hubble Volume simulations, we applied the FOF
identification algorithm to
a numerical simulation with box size $600$h$^{-1}$Mpc with
exactly the same cosmological parameters as the $\tau$CDM
simulation.  This ensures that this particular sample of
clusters is statistically comparable to those
obtained from the mock galaxy catalogue produced by the Durham group. 
\item Sample 2 comprises the results from the 
identification of groups in a mock catalogue that includes
a selection function, and where the position of particles is determined 
from redshifts, that is including particle 
peculiar velocities. We use a friends-of-friends
algorithm with different linking lengths in the directions parallel
and perpendicular to the line of sight.  This procedure emulates
the identification process used in the construction of the
group sample obtained from the Updated Zwicky Catalogue, which
is based on an algorithm described by Huchra \& Geller (1982).
In this case we use different linking lengths in
the directions perpendicular and parallel to the line of sight,
which vary linearly with distance,
$d_0=V/V_{fid} 0.229$h$^{-1}$Mpc and $v_0=V/V_{fid}350$km/s, where
$V$ is the distance, and $V_{fid}=7000$km/s.
These values correspond to a density contrast $\delta \rho/\rho \simeq 80$.
\item Sample 3 is obtained using the algorithm
presented by Lumsden et al.(1992), which is
used in identifying clusters from the COSMOS catalogue.
Following the prescription described by Lumsden et al. (1992),
we use an angular search radius $r_{c}=1.0$h$^{-1}$Mpc, and  
apply this procedure to the same mock catalogues as the second algorithm. 
We tested this identification algorithm by applying it to
the COSMOS galaxies, and comparing the outcome of the identification algorithm
to the positions of the EDCC clusters  (Lumsden et al. 1992).  We found out
that $\sim75 \%$ of the EDCC clusters were re-obtained by our identification
procedure.  We also applied this to a numerical simulation, and found that
only $\sim70\%$ of the clusters identified in the
simulation cube with a FOF algorithm were found
by our angular identifier.  More importantly, only $\sim60\%$ of the clusters
identified are real clusters.  These numbers illustrate
the degree of projection effects sample 3 is subject to, and
its study will serve as a test of how these effects
translate into the cluster correlation function in redshift-space.
\item Sample 4: 
We extract from sample 3 
the subset of clusters whose angular positions are coincident
(within the cluster identification radius) to those
in sample 2. This sample is
derived from a 2 dimensional catalogue and is confirmed to have a
redshift space identification at the same angular position.
Cluster distances are determined by using the galaxy members
identified from the angular position data.  This sample should
be less affected by projection effects than that obtained
from the angular data alone. However, cluster distances may include 
particles that are not physically related to any particular
cluster at all.  We find that
only $\sim50\%$ of the clusters from sample 3 are present in this sample
indicating that this procedure
provides a significant restriction.  We also notice that this fraction is slightly different than that resulting
from comparing the sample 3 clusters with those obtained from using
the FOF algorithm on a full simulation box. 

\end{itemize}
All the samples defined in this section include groups with a minimum
mass threshold.  This threshold is set so as to make the
space density of all the samples as close to that of sample 1 as possible.

In figure \ref{fig:pixel_2dconf}.  
we show the different results from the identification of
clusters of galaxies in the mock catalogue using angular and 3-D 
galaxy positions. This
figure shows the density of galaxies in the mock catalogue in pixels, and
clusters of galaxies in circles.  The gray scale of the pixels indicate
number of galaxies; the darker the pixel the higher is 
the number of galaxies
in it.  As can be seen, the circles enclose regions of high density of 
galaxies.  The different line styles with which the circles are drawn
correspond to the different algorithms used in the
identification of clusters.  The double-line circles correspond to clusters
from sample 3, that is, clusters identified using angular data.  In this
case, the radius of the circle corresponds to the radius out to 
which the counting of galaxies was done when measuring the cluster richness.
The thick line circles show clusters from sample 2, 
identified from 3-dimensional
positions affected by the peculiar velocities of the clusters.  In this  case,
the radius of the circle is proportional to the richness of the clusters.
By inspection to this figure, 
we can appreciate the large fraction of clusters from sample 3 
whose angular positions are not coincident with any
3-dimensional cluster within one angular search radius.

The aim of constructing the mock cluster samples presented here
is to understand and reproduce the results
obtained from observational cluster samples. In particular, 
we intend to establish the reason for the
large elongations along the line of sight present in the
measured correlation functions.   
The different behaviours of the correlation
functions found for the four mock samples are 
shown in figure \ref{fig:dist.sim}, where
we used clusters of galaxies
with distances measured using $10$ member galaxies.
The results from samples 1 and 2  show
the expected infall pattern.
This simply indicates
that the FOF algorithm still works well when
applied to a catalogue which incorporates a selection function and 
includes the effects of galaxy peculiar velocities.  
The infall is not so evident when inspecting the remaining
panels, which simply show the different degrees of projection effects
the different samples are subject to.  It can be seen that samples 3 and
4 show somewhat large elongations along the line of sight. Taking into
account that all panels show the results for the same value of $n_z=10$, this
is a clear signature of projection effects.  
The fact that sample 4 shows elongated contours along the line of sight
is indicative of substantial projection contamination.   Given that we have
constructed this sample by requiring an angular coincidence between 2-dim
and 3-dim identified clusters, such
anisotropies reflect the fact that there is a strong
contribution by other structures along the line of sight in producing
the observed distortion pattern.

In the following section we provide detailed descriptions of the characteristics
of each sample, analysing the differences in elongations arising
from using different $n_z$.

\subsection{Effects of observed number of redshifts}
\label{sec:sim.nz}

In order to make a thorough comparison between the mock cluster
samples and the observational correlation functions, we study in detail the
effects of using different $n_z$ in each of our mock samples.

%

\begin{figure*}
{\epsfxsize=16.truecm \epsfysize=8.truecm 
\epsfbox[31 313 567 561]{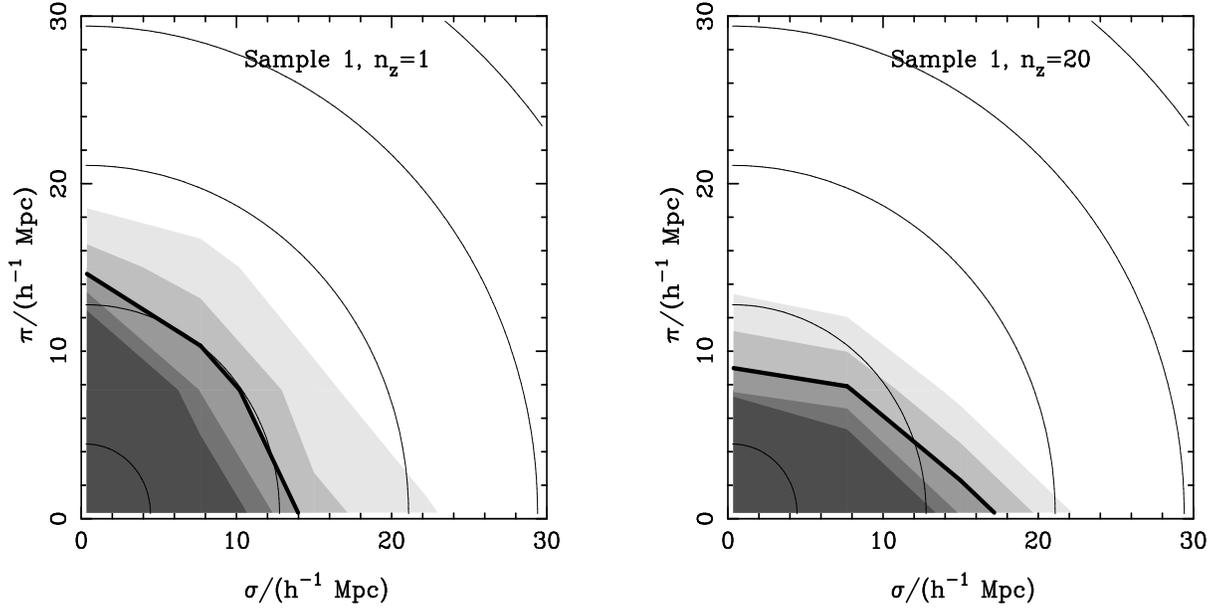}}
\caption{
The correlation function
contours for clusters identified using the FOF algorithm from the
simulation 
(mock sample 1).  The different panels correspond to 
different $n_z$:
$1$ redshift measurement (left) and $20$ (right)
member redshifts.
Shadings and line conventions are as in figure 1.
}
\label{fig:dist.sim.3d}
\end{figure*}

We first show the results from sample 1 which comprises
clusters of galaxies identified from the 3-dimensional
distribution of particles in the full numerical simulation.
Figure \ref{fig:dist.sim.3d} shows the correlation function
contours when the distances to the clusters in this sample have been
calculated using $1$ (left)
and $20$ (right panel) member redshifts.  As can be seen,
the differences between the panels are quite small.  The left
panel only shows slight evidence of an elongation along the line of sight.
This is an effect produced by the errors in the distance to these clusters
arising from the use of only one redshift to determine their distances.
By comparison between the left upper panel in figure \ref{fig:dist.sim}
and the right panel in figure \ref{fig:dist.sim.3d} (that is, Sample
1 with $n_z=10$ and $20$),
it can be seen that the use of a small $n_z=10$ is sufficient
to erase spurious elongations along the line of sight, and 
that considering more redshifts per cluster makes no significant difference. 
This result once more reflects the reliability of the FOF algorithm 
in finding clusters of galaxies in numerical simulations. 


\begin{figure*}
{\epsfxsize=16.truecm \epsfysize=8.truecm 
\epsfbox[31 313 567 561]{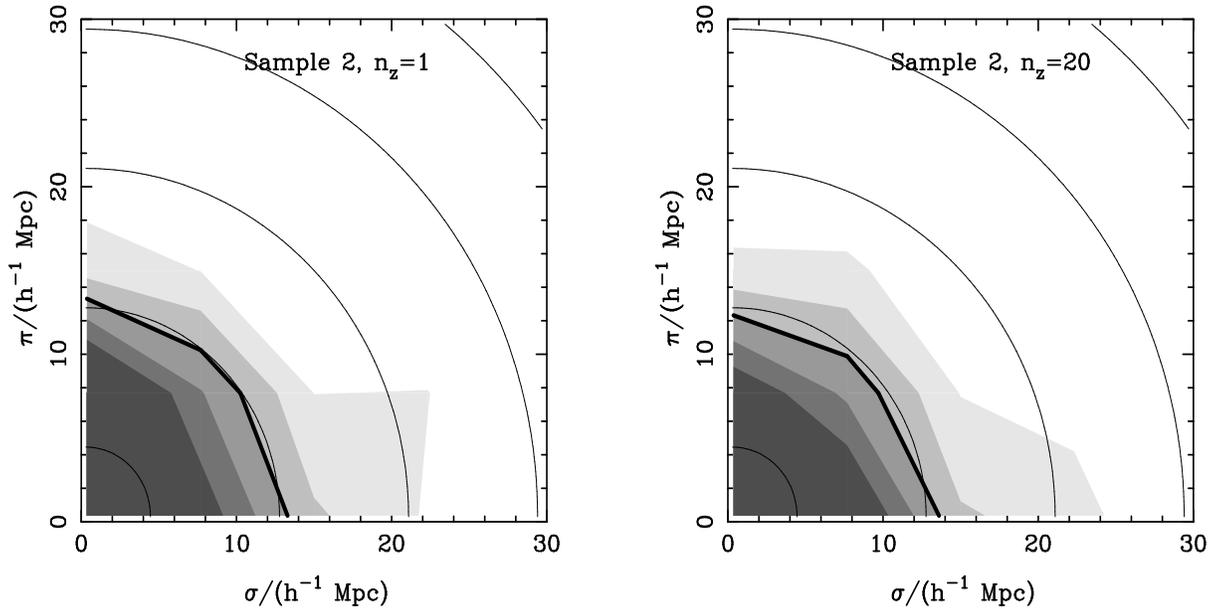}}
\caption{
Correlation function
contours for the cluster mock sample 2 which comprises clusters
identified using a modified version of the FOF algorithm, applicable to
a mock redshift survey, which includes peculiar velocities and a selection 
function.  We show the results for cluster distances obtained using
$1$ and $20$ redshifts in the left and right panels respectively.
Shadings and line conventions are as in figure 1.
}
\label{fig:dist.cole.3d}
\end{figure*}

We now investigate the effects of changing $n_z$ in sample 2, which consists of
clusters identified from a mock galaxy catalogue 
which incorporates a selection function. 
We show in figure \ref{fig:dist.cole.3d} the correlation function
contours for samples with distances obtained using
$1$ and $20$ redshifts in the left and  right hand side panels respectively.
As can be seen in this figure, the effects of using a small 
$n_z$ are not critical.  In the case of using $1$ galaxy, 
the infall pattern can still be seen, specially in
the correlation contours corresponding to $\xi(\sigma,\pi)<1$.  The inclusion
of further redshifts only makes a small change, although it can be seen that 
in this case, this signature is also visible for levels of higher
correlation function.  
When comparing with figure \ref{fig:dist.sim.3d}
one can notice a neat infall pattern in the clusters from sample 1
for high levels of correlation even
when $n_z=20$ (a very
clear signature for $\xi=1$ contour levels, and mildly visible 
for $\xi=1.4$).  The case of sample 2 is less ideal, since this infall pattern
can only be marginally seen for a correlation function contour level 
corresponding to $\xi=1$, which simply reflects the higher degree of difficulty
in identifying clusters from a galaxy catalogue which
incorporates a selection function.

It should be noted that sample 2 is representative 
of cluster or group catalogues identified
in galaxy redshift surveys, as is the case with UZC groups.  
Therefore,  
the infall of structures which is a signature of the 
hierarchical clustering obtained in the correlation function contours 
measured using the UZC groups sample (Padilla et al. 2001), 
can be assessed by the results of sample 2.
In other words, the study 
arising from sample 2 results in a correlation function which is not 
badly affected by projection effects, even for small $n_z$,  
which also indicates that the use of a small $n_z$ does not induce important 
elongations along the line of sight.  

\begin{figure*}
{\epsfxsize=15.truecm \epsfysize=15.truecm 
\epsfbox[31 22 567 561]{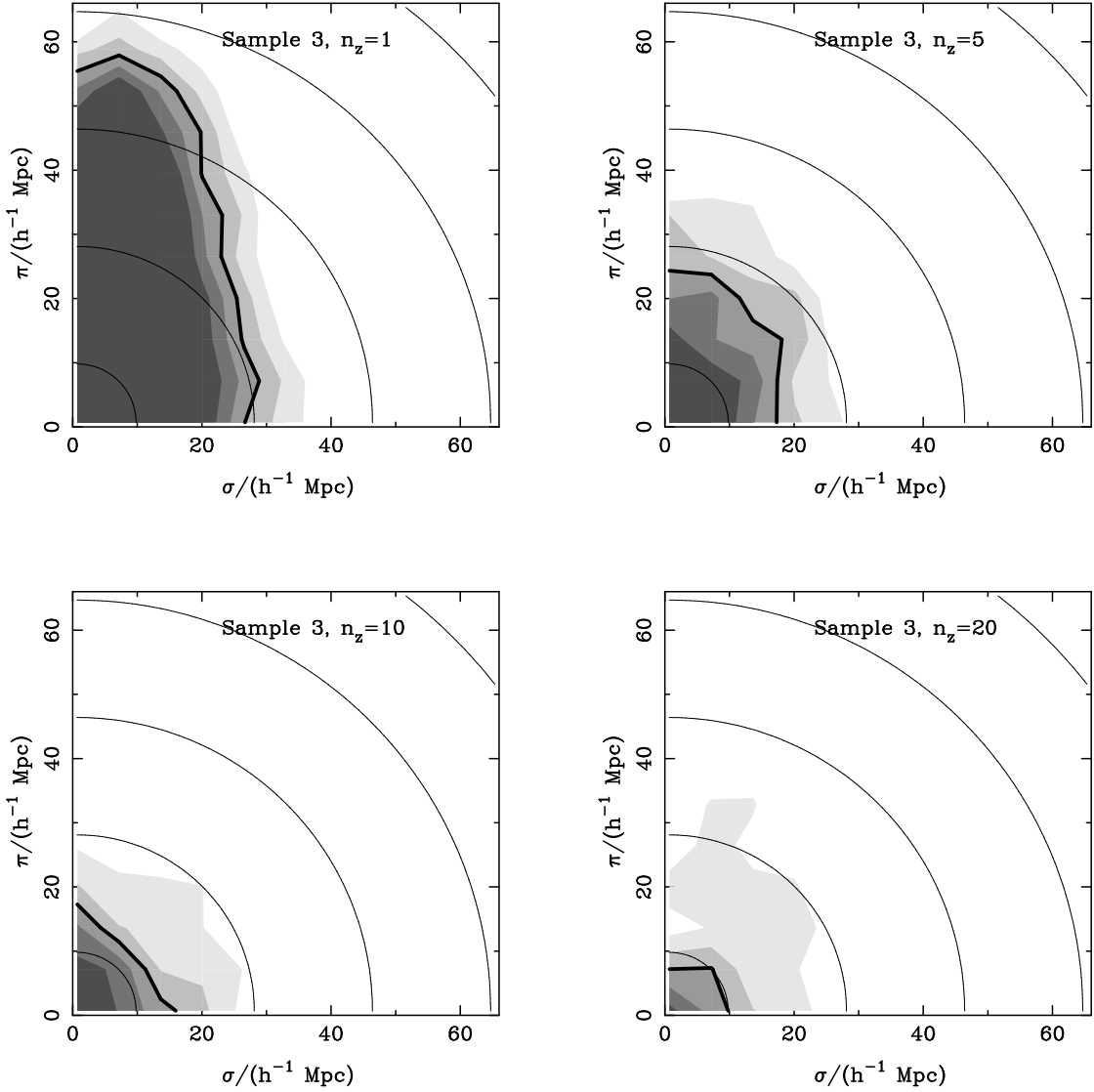}}
\caption{
Correlation function isopleths obtained from clusters identified
from angular positions drawn from the numerical simulation 
(mock cluster sample 3),
with $1$, $5$, $10$, and $20$ individual redshift measurements in panels
upper left, upper right, lower left, and lower right respectively.
Shadings and line conventions are as in figure 1.
}
\label{fig:dist.cole.2d}
\end{figure*}

We next  consider a mock cluster catalogue drawn from an angular 
galaxy catalogue (sample 3).
Here the distances to each cluster were
obtained using the redshifts of galaxies identified as cluster members 
using angular data.  The correlation
function of these catalogues are shown in figure \ref{fig:dist.cole.2d}, 
where we show the results when using $1$, $5$, $10$, and $20$ galaxy 
redshifts in determining cluster distances in the upper left, 
upper right, lower left and lower right panels respectively.

The results in this figure 
show significant differences in the elongations for the different
samples.
In the case where only $1$ galaxy redshift is used to assess the cluster
distance, the elongation along the line of sight is severe, and
not entirely different to what was found for Abell clusters with distances
estimated using less than $10$ galaxy redshifts (see for instance
Bahcall, Soneira \& Burgett, 1986).  The use of $5$ galaxies already makes a 
marked difference but would still not be sufficient for an infall pattern
to be found.  For comparison, the number of Abell clusters
with $n_z \geq 5$ is small, about 20 \% 
of the total sample.  The first, faint traces of an infall
can be seen when considering $10$
galaxy redshift measurements, and only for small values of $\xi(\sigma,\pi)<1$,
which are difficult to obtain from an observational sample without large
contributions from noise.

The main sources of elongation cannot be considered to be the error in the
cluster position arising from using a small $n_z$.
This is so since, for instance, 
we have already shown in figure \ref{fig:dist.cole.3d} that
the use of a very small number of galaxies is still not enough to
erase the infall pattern.  It is more likely that the elongation
is produced by the inclusion of non member galaxies
in the determination of cluster distances, the worst scenario being
that in which only a fraction of the identified clusters correspond
to physically bound galaxies, and the remainder being constructions of
projection effects that cause the spreading of galaxies along the line of sight
to appear as coherent structures, as only angular data is available.

We now consider what are the expected effects of the number
of redshift measurements in samples of clusters identified in 
two dimensions with angular positions coincident with 3-dimensional
identification.

\begin{figure*}
{\epsfxsize=15.truecm \epsfysize=15.truecm 
\epsfbox[31 22 567 561]{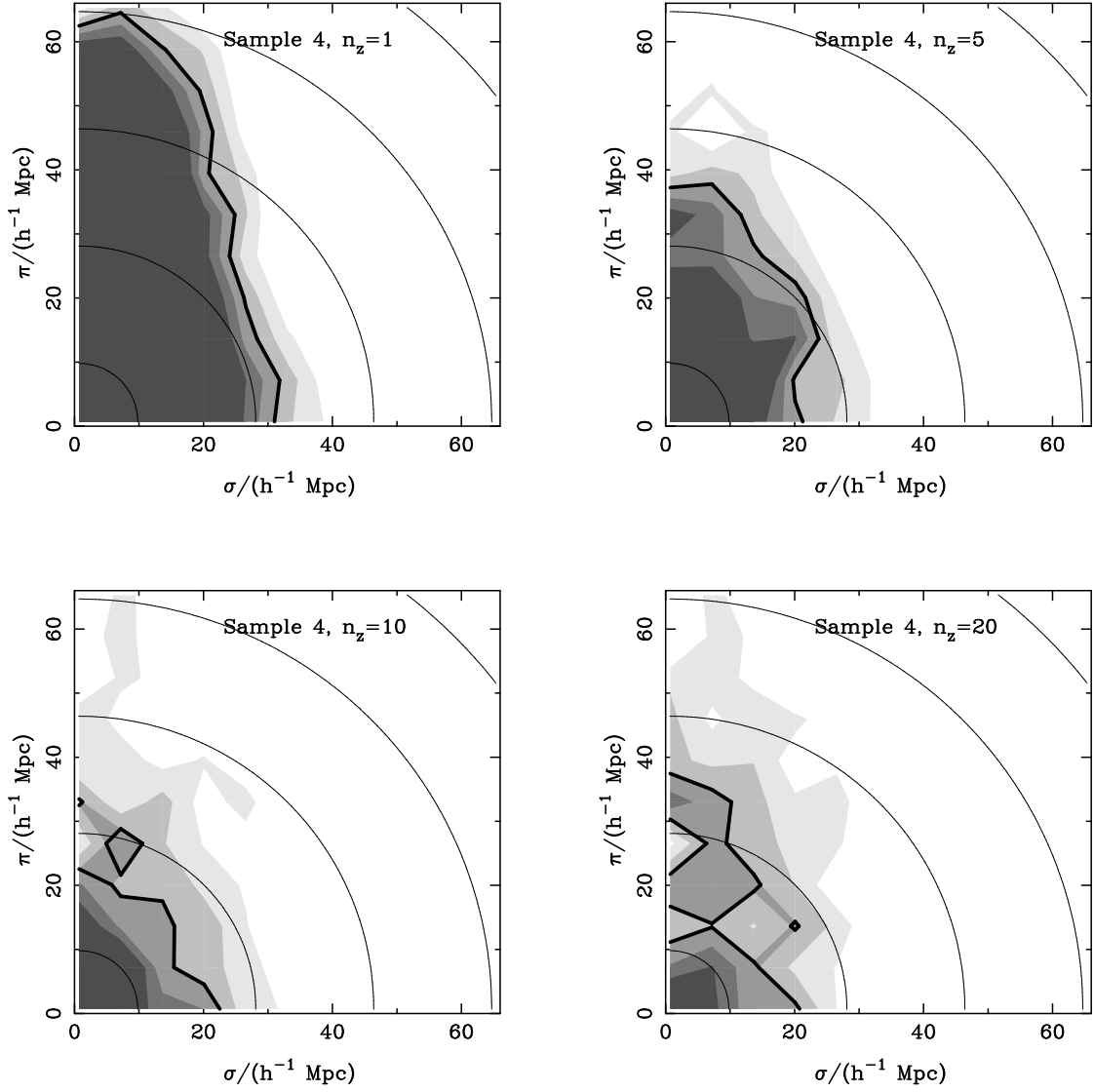}}
\caption{
Correlation function results from clusters identified using
mock angular data using 2D, and whose angular positions coincide
with those of real clusters in the simulation
(mock sample 4).  Cluster
distances were obtained using $1$, $5$, $10$ and $20$ redshifts
(panels upper left, upper right, lower left, and 
lower right respectively).
Shadings and line conventions are as in figure 1.
}
\label{fig:dist.cole.2dconf}
\end{figure*}

Figure \ref{fig:dist.cole.2dconf} shows the correlation function contours
for clusters in mock sample 4.  The different panels correspond
to different $n_z$,
in the same order as the previous figure.

As can be seen in figure \ref{fig:dist.cole.2dconf}, 
the projection effects in 
sample 3 are still present, even after the clusters have 
coincident angular positions with
a cluster sample identified in 3D.  
The elongations in $\xi(\sigma,\pi)$
are quite severe, and reflect the contamination of foreground and background
structures, which introduces biases in the cluster distances.

\subsection{Effects of projected radius in the identification algorithm}

The algorithms used to identify clusters from angular data depend 
quite sensitively on the search radius $r_c$ used to find
member candidates.  In the case of
Abell clusters this corresponds to 
$r_{c}=1.5$h$^{-1}$Mpc. 

\begin{figure*}
{\epsfxsize=17.5truecm \epsfysize=5.3truecm 
\epsfbox[31 22 856 271]{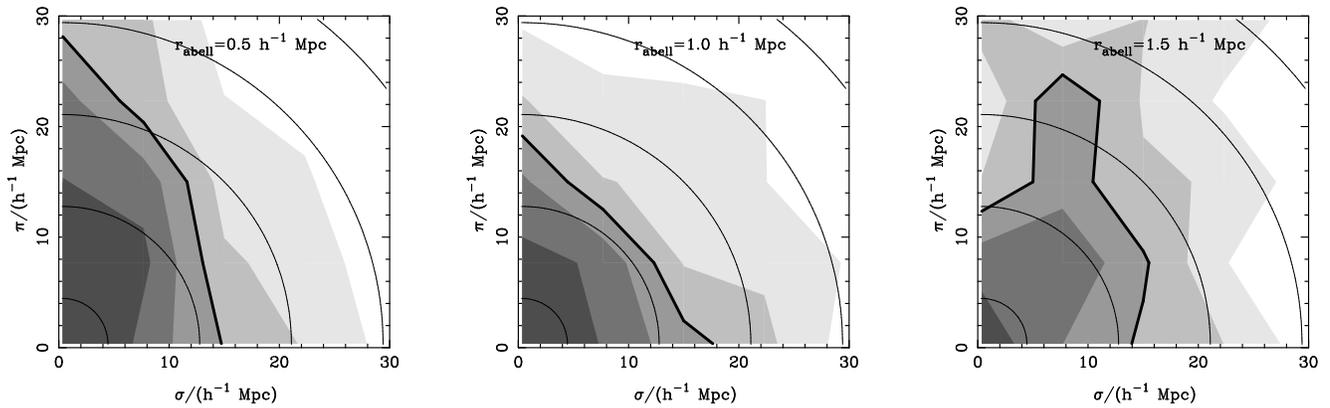}}
\caption{
Correlation function
contours for mock clusters identified from angular data,
using $r_{c}=0.5$, $1.0$ and
$1.5$h$^{-1}$Mpc in the left, middle and right hand side panels
respectively. 
Shadings and line conventions are as in previous figures.
}
\label{fig:dist.cole.2d.abell}
\end{figure*}

A possible cause of projection effects becomes clear
when one considers that the  density of galaxies 
diminishes as a function of
cluster radius.  This means that in the outskirts of the cluster, 
the angular population of galaxies will be composed of ever larger
fractions of foreground and background galaxies.
Therefore it
is increasingly probable to consider as member galaxies, objects 
that are not
physically bound to the cluster.  We investigate this
effect by using different cluster search radius in our 2-dimensional
cluster finding algorithms.  In figure
\ref{fig:dist.cole.2d.abell}, we show the 
correlation function
contours for clusters identified using $r_{c}=0.5$, $1.0$ and
$1.5$h$^{-1}$Mpc in the left, middle and right panels
respectively.  In all these cases, 
$n_z = 10$, which ensures us that any elongations
observed in the $\xi(\sigma,\pi)$ isopleths will be mainly
produced by projection effects arising from the choice of radius.

\begin{figure}
{\epsfxsize=8.5truecm \epsfysize=12.75truecm 
\epsfbox[ -20 1 370 560]{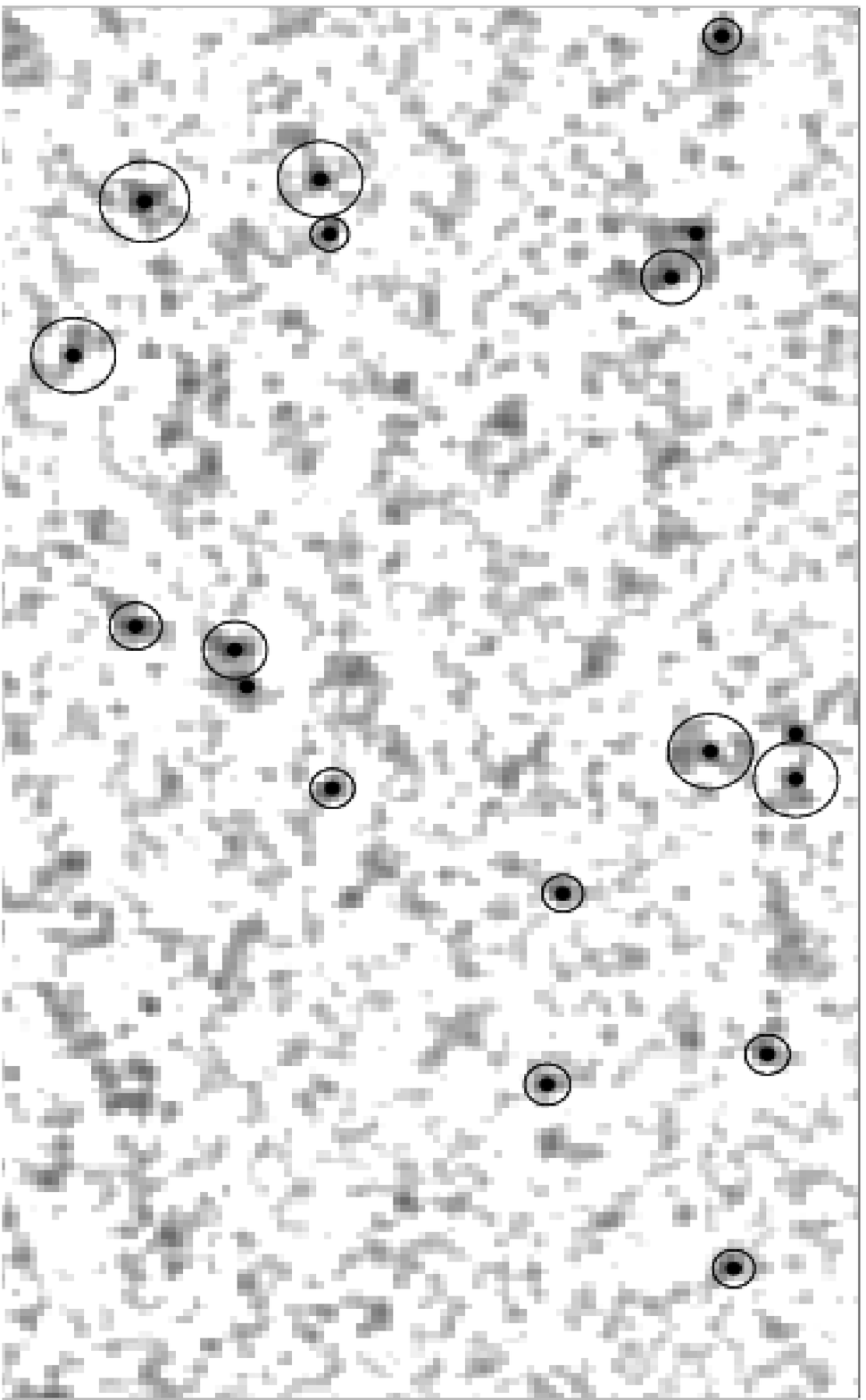}}
\caption{
Clusters in mock sample 3 identified with different
cluster search radius.  The density
of galaxies is shown as a grey-scale pixelmap.
The circles show clusters identified using $r_{c}=1.0$h$^{-1}$ Mpc,
and the dots, those identified using $r_{c}=0.5$h$^{-1}$ Mpc.
This figure shows a sub-set of the
pixelised mock galaxy catalogue and is shown 
in an equal area projection centred at
$\alpha=1^h$ and $\delta=-45^o$, with radius $r_{eqa}=160^o$.  The
number of pixels per side is $n_{pix}=2900$, and was set in order
to obtain a mean number of galaxies $n=1$ per pixel for 
a magnitude limit $m_{gx}<20.0$.  The pixelmap sub-set
size is $110{\rm x}200$ pixels.
}
\label{fig:pixel_diffabell}
\end{figure}

As can be seen, the use of a small search radius is not
entirely convenient as it produces spurious clusters, probably
from confusing satellite haloes with proper clusters.  This
is in agreement with the elongated iso-correlation contours in 
the left panel, and also with the small amplitude of the
correlation length $\sigma_0$, that is $\xi(\sigma_0,\pi=0)=1$, 
and is supported by   
figure \ref{fig:pixel_diffabell}, which
shows the 
clusters of galaxies identified
using $r_{c}=1.0$h$^{-1}$Mpc (circles), and those
identified using $r_{c}=0.5$h$^{-1}$Mpc (dots);  
the gray scale of the pixels indicate number of galaxies, and
the radii of the circles correspond to the radius out to 
which the counting of galaxies was done when finding the cluster richness.
As can be seen 
in three cases in figure \ref{fig:pixel_diffabell}, 
the identification using a small search radius
produces pairs of clusters whereas the use of a larger search radius 
produces a single cluster.  The satellite cluster identified with a small
search radius corresponds to a lower peak in the density of projected galaxies
in all cases; also, the richness of a satellite is smaller than the
main cluster in general, 
confirming the hypothesis of low mass satellites 
stated earlier in this paragraph.

The other values considered, $r_{c}=1.0$ and $1.5$h$^{-1}$Mpc, 
produce similar results, both largely free from projection effects, 
and with a larger correlation length $\sigma_0$ (figure 
\ref{fig:dist.cole.2d.abell}).  There is a slight
difference in the noise level in these samples, the correlation
levels in the $r_{c}=1.5$h$^{-1}$Mpc sample being
less smooth than the smaller
$r_{c}$ case.  This is not a severe difference, but is enough to support 
our choice of $r_{c}=1.0$h$^{-1}$Mpc used throughout this paper.

\subsection{Relative velocities and
correlation lengths in mock cluster samples}

\label{sec:sim.wdnz}

\begin{figure*}
\epsfxsize=15.truecm \epsfysize=15.truecm 
\epsfbox[30 275 516 715]{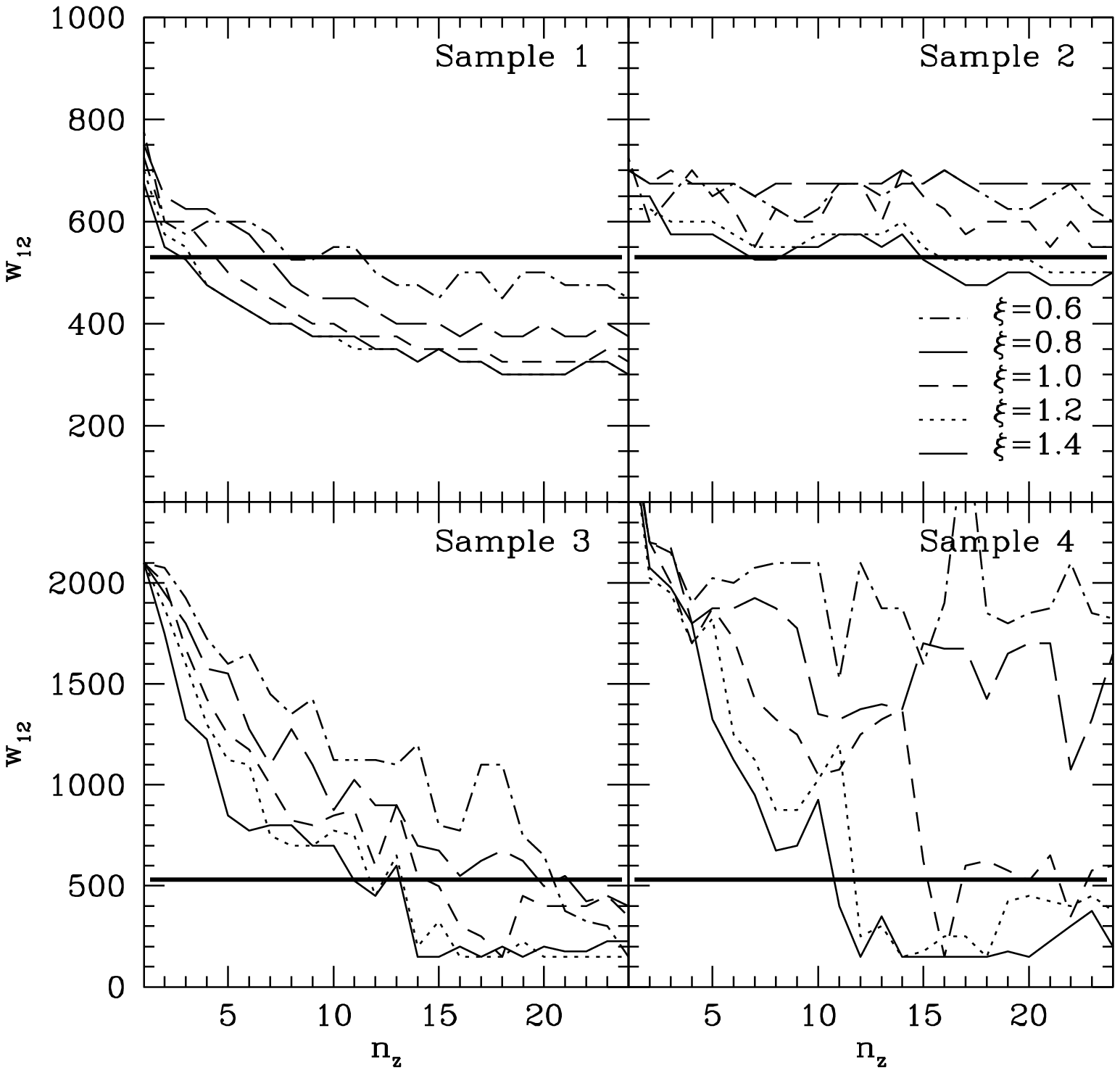}
\caption{
Pairwise velocities obtained for the different samples
as indicated in the legend of each panel.  The different
thin lines correspond to different values of $\xi(\sigma,\pi)$.
The thick solid lines show the mean pairwise velocity dispersion
for the full sample of clusters in the 3-dimensional
simulation box.
}
\label{fig:wdnz}
\end{figure*}

In this section we present determinations  
of relative velocities, redshift-space correlation lengths and
bias factor from the $\xi(\sigma,\pi)$ anisotropies.
Specifically,  we study the dependence of the relative pairwise
peculiar velocities obtained using eq. \ref{eq:chi} for
the different mock samples as a function of $n_z$.
We recall that the cosmological parameters of 
the simulation from which the mock catalogues were extracted
($\Omega_m=1$, $\Omega_{\Lambda}=0.0$,
$\Gamma=0.21$ and $\sigma_8=0.6$) are those used in the calculations.

Figure \ref{fig:wdnz} shows the pairwise velocities found
in the different samples (sample 1 in upper left panel, 
sample 2 in the upper right,
sample 3 in the lower left, and sample 4 in the lower right
panel).  The different thin
lines correspond to results using different values of $\xi(\sigma,\pi)$.
The thick solid line shows the mean pairwise velocity
dispersion from the clusters identified in three dimensions
in the simulation box.

This plot shows the great range of results which can be obtained
from the contours in the correlation function if the mean $n_z$ 
used for the clusters in the sample is not known.  
It is remarkable the fact that the relative velocities found using
clusters from sample 2 which were obtained in redshift space 
are in good agreement
with the actual value measured from the full simulation.

The disagreement between the value of true relative cluster
velocities with the measurements from
the redshift-space correlations 
is most severe when the sample has been identified
using angular data.  It is also interesting that the
velocities obtained from sample 4 (clusters identified from 2 dimensions
with angular confirmation from inspection over the 3-dimensionally
identified sample) are in some cases still very high even when 
their distances are obtained
using a large $n_z$.  It should be noted though, that
the velocities favoured by low noise contour levels, 
$\xi \geq 1$, suggest
small cluster relative velocities.  
The problem of a wide range of predictions for $w_{12}$ 
is not present in the case of clusters
in sample 3, in part
due to the larger number (almost a factor of 2) of clusters in this sample.  
The results obtained from 
different correlation levels are more compatible, 
showing values differing by
less than $1000$ km/s among the different correlation levels. 
The observed behaviour of the relative velocities does not  
favour sample 4 over sample 3, indicating that the confirmation
by angular coincidence with real clusters
can not remedy the systematic problems of samples of clusters identified in two 
dimensions.

\begin{figure}
\epsfxsize=8.truecm \epsfysize=8.truecm 
\epsfbox[24 155 576 700]{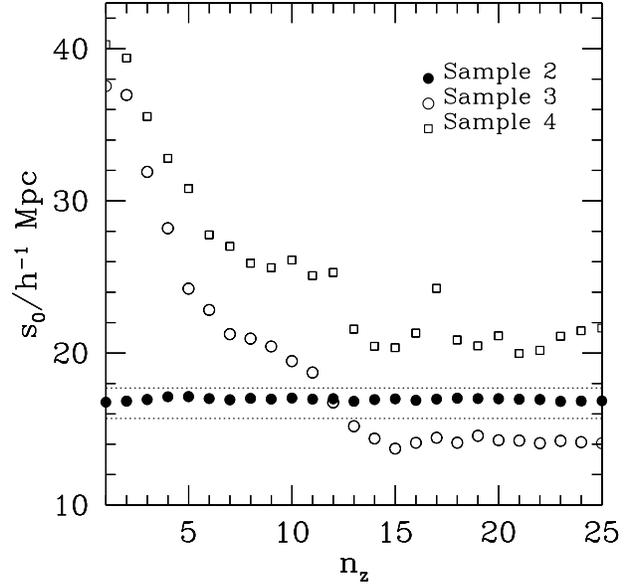}
\caption{
Values of the redshift-space correlation length $s_0$
as a function of $n_z$.  The symbols in these 
panels correspond to the different samples and are indicated in the figure.
The dotted lines show the range of acceptable values as measured
from the full simulation.
}
\label{fig:s0}
\end{figure}

This problem can be dug into a little further by studying
the cluster redshift-space correlation lengths, which also show great 
variation with $n_z$.  We show these quantities in figure \ref{fig:s0}, 
where marked differences between 
samples of clusters identified from angular and 3-D data can be seen.
Both samples obtained using 3-D information show a steady value
of $s_0$, without any significant variation with $n_z$.
The first important result from this figure is that
the results from clusters in sample 2 are in excellent
agreement with the true values.  
We notice that the results
for sample 1 are indistinguishable from those of sample 2
and are not shown for clarity.

An interesting feature in figure \ref{fig:s0}, is the difference between
confirmed and unconfirmed 2-D clusters, which also show controversial
results in figure \ref{fig:wdnz}. 
The observed correlation
length of clusters in sample 4
is higher than that of redshift space identified clusters in sample 2.
Furthermore, this correlation length is also  higher than that of the clusters
in sample 3, which is expected to be more contaminated by
foreground/background galaxies.  This differences could be explained
if several spurious clusters in sample 3, were associations of a small number
of galaxies along the line of sight.
The higher correlation amplitude derived for sample 4 clusters with respect
to those in sample 2, could indicate
that the identification of clusters using angular data is biased
towards massive systems.  This follows from the
larger values of $s_0$ found from the sample of confirmed clusters.
The inclusion of spurious small associations
of galaxies in sample 3, would make the value of $s_0$ in this case 
smaller than those obtained for clusters in sample 2.
We have also confirmed this by inspection to the mean cluster
mass of samples 2 and 4.

These findings are in agreement with the results of Miller
et al. (1999), who showed that by restricting a sample
of Abell clusters to
high richnesses the degree of
projection effects is significantly reduced.
The difference in correlation lengths also explains 
the marginally larger cluster relative velocities
found in sample 4 compared with sample 3, at least when considering
high levels of correlation, where the noise is not
playing an important role in the results from sample 4.
Equation \ref{eq:streaming} shows that the streaming velocities
of clusters are proportional to their correlation functions.
This means that the clusters in sample 4 have larger streaming
velocities than those in sample 3.  Taking into account that
the elongations along the line of sight seen in figure 
\ref{fig:dist.cole.2dconf} are similar for both samples, it is clear
that the results from equation \ref{eq:chi} will yield larger
values of cluster relative velocities for the sample 4
than for sample 3.

\begin{figure}
\epsfxsize=8.truecm \epsfysize=8.truecm 
\epsfbox[54 195 576 720]{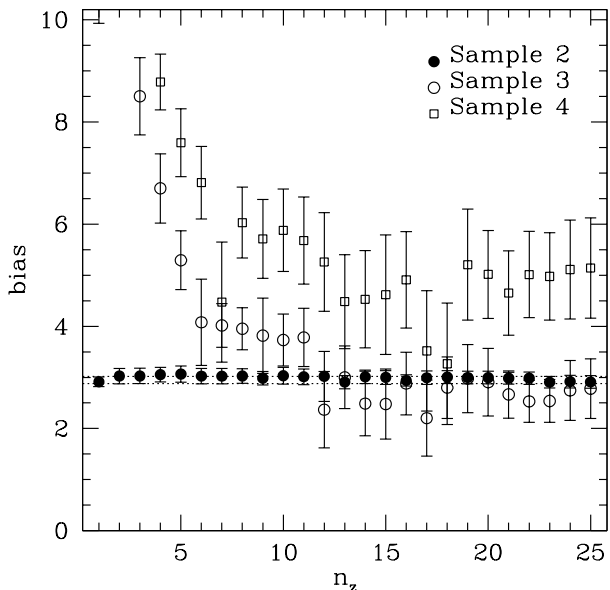}
\caption{
Values of the bias parameter obtained from minimising equation 
\ref{eq:chi} as a function of $n_z$.  The symbols in these 
panels correspond to the different samples and are indicated in the figure.
The dotted lines show the acceptable range of bias parameters for
the number density of clusters in the mock samples obtained
using the Sheth, Mo \& Tormen (2001) mass function.
}
\label{fig:bias.sim}
\end{figure}

The results for the values of the bias parameter obtained from 
minimising equation \ref{eq:chi} as a function of $n_z$ show a behaviour
similar to the results from the redshift-space
correlation length.  Figure \ref{fig:bias.sim} shows
the resulting bias parameter for the different
mock samples (the symbols are indicated in the figure).
The dotted lines show the acceptable range of 
effective bias parameters from the simulation,
obtained using the Sheth, Mo \& Tormen (2001) mass function, and
the number density of clusters in the mock samples along with 
its uncertainty. 

As it can be seen, the bias found for the clusters in mock sample 2
is within the
acceptable values for the CDM effective bias.  This result
holds for any value of $n_z$.  As in the previous
figure, the results for sample 1 are consistent
with the true values, and are not shown in order to preserve clarity.
The result from sample 3 is in agreement for roughly $n_z>10$.  
Sample 4 shows significantly larger values of  bias and correlation
length, suggesting that this sample comprises a high mass
cluster population in agreement with the results found earlier in this
section.

\section{Conclusions}

We have analysed the consequences of
different effects on the distortion of 
the correlation function of clusters in redshift space
for different mock samples of galaxy clusters. 
We take into account
the pairwise velocity distribution of galaxy systems, coherent 
bulk motions, as well as redshift errors, and different 
cluster identification systematics.
The mock cluster samples were extracted from
numerical  simulations, following algorithms that closely match
the procedures used to identify the clusters in observational samples. 
We find that the correlation function of galaxy systems is influenced 
by wrongly assigned cluster distances due to a small number
of galaxy redshift measurements in the fields of clusters. 
However, the most important effect is that associated with the cluster 
identification procedure from two dimensional surveys. Due to these effects, 
the estimated mean pairwise 
velocity dispersion can be an order of magnitude larger than
the actual value. 

This
is consistent with the observed anisotropy of the correlation function contours
of UZC groups, which shows the flattening produced by this infall
motion (Padilla et al. 2001).  
For samples of clusters identified in mock galaxy redshift surveys
we find an infalling pattern even in the case where
the distances to groups or clusters of galaxies are obtained using a small
number of member redshifts. 

\begin{figure}
\epsfxsize=7.5truecm \epsfysize=7.5truecm 
\epsfbox[40 30 550 550]{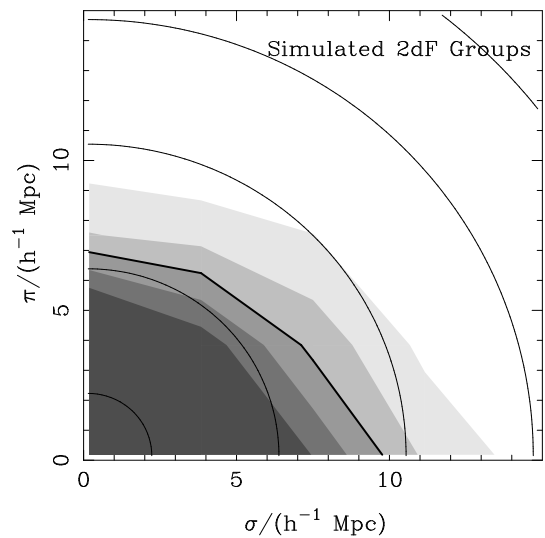}
\caption{
Correlation function results from mock 2dF100k release galaxy groups
constructed by Merch\'an \& Zandivarez (2002).
Shadings and line conventions are as in figure 1.
}
\label{fig:dist.2dfmock}
\end{figure}

In order to provide predictions for new generation cluster catalogues 
we have also analysed a mock sample of groups resembling
those obtained by Merch\'an \& Zandivarez (2002) from the 2dF100K release 
galaxy redshift survey. This
mock catalogue is subject to the same complicated angular mask as the observations and serves as a further test of the reliability of the information that can be
obtained from observational cluster and group catalogues. The results 
shown in figure \ref{fig:dist.2dfmock}
are similar to those of Sample 2, and indicate 
that the angular mask does not affect severely the results so that 
consistent values of $w_{12}$ and $s_0$ can be derived.  

In order to explore the projection effects afflicting cluster samples
identified in two dimensions
we analysed different mock cluster samples
identified from mock angular catalogues using different cluster search radius.
Our findings favour the use of smaller radius than that used in
the identification of Abell clusters.  The use of a too small 
value produces a sample increasingly affected by projection
effects and the inclusion of smaller groups of galaxies which are probably
cluster satellite haloes.

We notice that 2-dim identified clusters whose angular positions
coincide with a true 3-dim cluster show a strongly distorted clustering
pattern.  This fact shows the large influence of groups and clusters
along the line of sight in the observed elongation.  Contamination 
is therefore mainly arising from structures present in the filamentary
large scale distribution.  These results indicate the difficulties
of obtaining unbiased cluster samples even when using accurate photometric
data such as present in new and forthcoming surveys.  Redshift information
is crucial for constructing cluster samples which lack significant
projection effects, since galaxy colours alone can remove foreground
and background contamination but not distinguish membership to clusters at
relatively smaller separations along the line of sight. 

\section*{Acknowledgments}
This work  was supported in part by CONICET, Argentina,
and the PPARC rolling grant at the University of Durham.  
DGL acknowledges support from the John Simon Guggenheim Memorial Foundation.
We thank the Referee for invaluable comments and advice
which greatly improved the previous version of the paper.
We have benefited from helpful discussions with Carlton Baugh.
We acknowledge the Durham Extragalactic Astronomy Group and the 
Virgo Consortium for making the Hubble Volume 
simulation mock catalogues available.

\end{document}